\documentclass[12pt]{article}
\usepackage{epsfig}
\usepackage{amssymb,amsfonts}
\newcommand{\beq}{\begin{equation}}
\newcommand{\eeq}{\end{equation}}
\newcommand{\bea}{\begin{eqnarray}}
\newcommand{\eea}{\end{eqnarray}}

\begin{document}

\title{\bf Stability of strange quark matter: model dependence}

\author{
W.M. Alberico,
C. Ratti
\vspace*{0.3cm}\\
\begin{tabular}{c}
{\it Dipartimento di Fisica Teorica, Universit\`a di Torino}\\
{\it and INFN, Sezione di Torino, 
Via P. Giuria 1, 10125 Torino, Italy}
\end{tabular}
}

\maketitle
\begin{abstract}
The minimum energy per baryon number of strange quark matter is studied, as a 
function of the strangeness fraction, in the MIT bag model and in two 
different versions of the Color Dielectric Model: a comparison is made with 
the hyperon masses having the same strangeness fraction, and coherently 
calculated within both models. Calculations are carried out in mean field 
approximation, with one gluon exchange corrections. The results allow to 
discuss the model dependence of the stability of strangelets: they can be 
stable in the MIT bag model and in the double minimum version of the Color 
Dielectric Model, while the single minimum version of the Color Dielectric 
Model excludes this possibility.   
\end{abstract}

\section{Introduction}
The production of strange quark matter and/or hypermatter in central heavy ion 
collisions has been suggested long ago, either in the form of 
multi-hypernuclear objects (strange hadronic matter) 
\cite{schaffner92,schaffner93,schaffner94}, or strangelets (strange multiquark
 droplets) \cite{chin79,liu84,greiner87,greiner88,greiner91}. The formation of
 the latter would be rather appealing, since it would be an unambiguous 
signature that a deconfined, strangeness rich state of quark gluon plasma has 
been created during the reaction. The investigation of the strangelet
 stability is therefore of primary importance for their detection in heavy ion
 experiments.

The idea is that, even if no strangeness is present in the initial state of 
the collision, and no {\it net} strangeness is expected after the reaction, 
nevertheless a large number of $s\bar{s}$ pairs can be produced in a single 
central event; the antiquarks $\bar{s}$ are then able to rapidly combine
with the abundantly available $u$ and $d$ quarks to form antikaons that 
immediately leave the fireball region, which becomes strangeness rich matter. 
The hadronization process is then of fundamental importance: the
 copious formation of strange particles cannot be considered as a reliable 
signature of QGP formation, since kaons and hyperons can be produced in 
hadronic reactions as well \cite{dover93}.
If, on the contrary, after the formation of the deconfined plasma, this 
strangeness rich matter could coalesce into colorless multiquark states, the 
so-called strangelets, this would be an unambiguous signature of QGP 
formation; this process might be favoured by a rapid QGP cooling due to the 
prompt anti-kaon (and also pion) emission from the surface of the fireball.

Up to now, the stability of strangelets has been only investigated within the 
MIT bag model \cite{fahri84}, also including ${\mathcal{O}} (\alpha_s)$ 
corrections to the properties of bulk strange matter: according to this 
pioneering work, heavy, slightly positively charged, 
strangelets could be more stable than ordinary nuclei.
 A detailed calculation of strangelet properties within the MIT bag
model, including shell effects and all the hadronic decay channels has been
performed by J. Schaffner {\it et al.}~\cite{schaffner97}: a valley of 
stability clearly appears for $A_B= 5\div 16$ with charge fraction $Z/A$
between 0 and $-0.5$. On the other hand, strangelets having a larger mass 
should be positively charged according to the results of Ref.~\cite{madsen00}.

In ref.~\cite{alberico02} the authors confront the predictions about the 
stability of strangelets within the MIT bag model and the Color Dielectric 
Model (CDM): the equilibrium energy of the strange matter is compared with 
the masses of hyperons having the same strangeness fraction, and 
coherently calculated within both models.
 The main goal is to find out whether and to which extent 
the stability of strange matter and/or strangelets depends on the model 
employed to describe the confined system of quarks. The present contribution
is largely based on the results of ref.~\cite{alberico02}, with a special 
focus on the model dependence of the strangelets stability.
We consider homogeneous quark matter made up of $u$, $d$ and $s$ quarks, 
without imposing chemical equilibrium on the density of the strange quarks.
Rather, we  assume that there exists in the system  a definite strange 
fraction $R_s=\rho_s/\rho$, $\rho$ being the total baryon density of quarks 
and $\rho_s$ the baryonic density of strange quarks. This is coherent with the
 hypothesis that, during a high energy collision between heavy ions, 
this state of matter, if formed at all, can only survive for a very short time,
 so that it has no time to reach $\beta$ equilibrium; hence the minimal energy
 per baryon number can be studied as a function of the strange fraction $R_s$.
We also consider the effect induced by the introduction of perturbative 
gluons in both models. Since electromagnetic interaction has been neglected, 
the minimum of the energy corresponds to an equal number
 of $u$ and $d$ quarks.

We consider, for simplicity, an infinite and homogeneous system, but
strangelets are indeed finite objects, and therefore  one should remember that 
 the energy of the infinite system appears to be a lower
limit with respect to the envelop of  strangelet energies versus 
strangeness fraction: the latter was nicely illustrated by Schaffner {\it et 
al.}~\cite{schaffner97} calculating the strangelet masses within the MIT bag 
model with shell mode filling. We simply recall that
 surface effects, which we do not consider, would increase the energy curves
of bulk matter, typically of 50-100~MeV: hence, if hyperons should 
turn out to be more stable than strange matter, then this would exclude also
the stability of strangelets. If, on the contrary, strange matter is more 
stable, then this provides only an indication in favour of stable strangelets
unless the mass gap between the two states is large enough.

\section{Strangelets in the MIT bag model}
\subsection{MIT bag without gluons}
We consider first the simplest version of the MIT bag, not including one gluon 
exchange corrections; therefore the model has only two parameters: the vacuum 
pressure $B$ and the strange quark mass $m_s$. In order to discuss the various 
possible scenarios, we have used a wide range for both parameters: 
\begin{eqnarray}
B&=&60, 100, 150~\mathrm {MeV/fm^3};\\
m_s&=&100, 200, 300~\mathrm {MeV}.
\nonumber
\end{eqnarray}
The single flavor contribution to the energy density of the system is given by:
\begin{equation}
\epsilon_f=6\int\frac{d\vec k}{\left(2\pi\right)^3}E_f\left(k\right)\theta
\left(k_{F_f}-k\right),
\end{equation}
where $E_f(k)=\sqrt{k^2+m_f^2}$ and $k_{F_f}$ is the Fermi momentum of flavor
$f$. It can be analytically expressed, for $u$, $d$ (massless) and $s$ quarks,
 respectively:
\begin{eqnarray}
\epsilon_{u,d}&=& \frac{3}{\left(2\pi\right)^2}\,k_{F_{u,d}}^4\,,\\
\epsilon_s&=&\frac{3}{8\pi^2}
\left[m_s^4\ln\left(\frac{m_s}{k_{F_s}+\sqrt{k_{F_s}^2+m_s^2}}\right)
+k_{F_s}\sqrt{k_{F_s}^2+m_s^2}\left(2k_{F_s}^2+m_s^2\right)\right]\,.
\nonumber
\end{eqnarray}
The total energy density of our system turns then out to be:
\begin{equation}
\epsilon_{tot}=2\epsilon_{u,d}+\epsilon_s+B \,.
\end{equation}
The dependence of the above formula on $R_s$ and $\rho$ can be easily found 
by recalling the following relations for the various Fermi momenta:
\begin{eqnarray}
\rho_s&=&R_s\rho\\
\nonumber
k_{F_s}&=&\left(3\pi^2\rho_s\right)^{1/3}\\
\nonumber
k_{F_{u,d}}&=&\left(\frac{3\pi^2}{2}\rho\left(1-R_s\right)\right)^{1/3},
\nonumber
\end{eqnarray}
\begin{figure}[b]
\includegraphics[width=\textwidth]{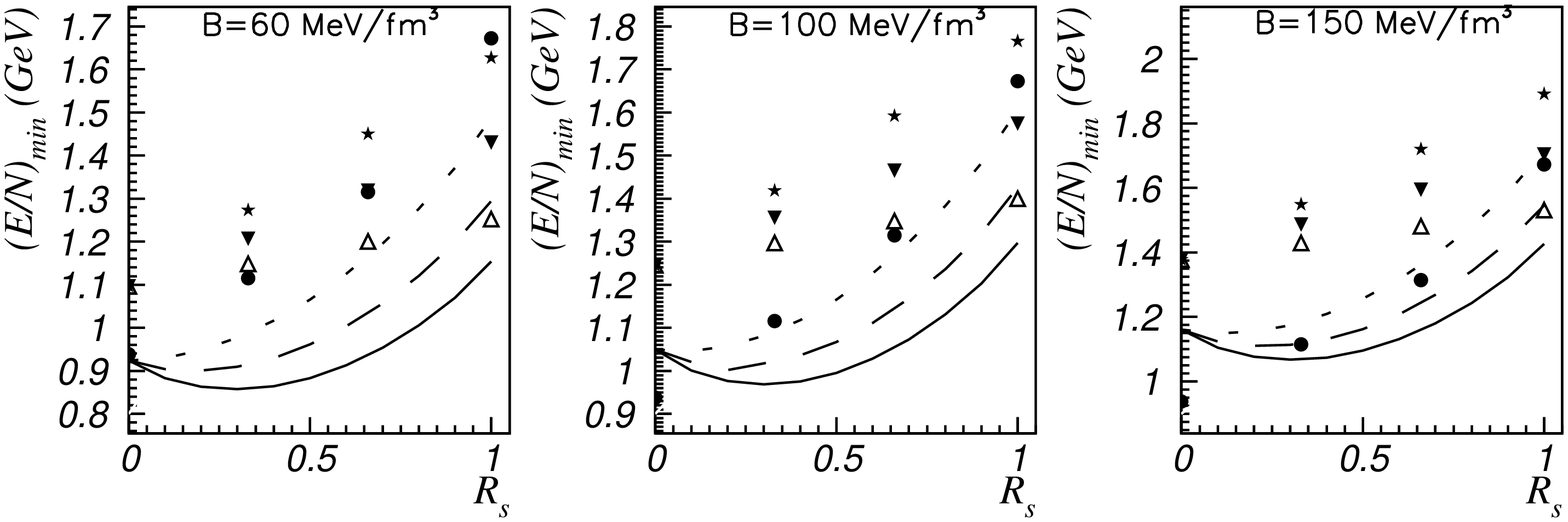}
\caption{Minimal energy per baryon number in the MIT bag model, 
as a function of the strangeness
 fraction $R_s=\rho_s/\rho$, for various values of the 
model parameters. The continuous line corresponds to $m_s=100$~MeV, the 
dashed line to $m_s=200$~MeV and the dotted line to $m_s=300$~MeV. Full circles correspond to experimental masses, the other points to the masses 
evaluated in the model, with $m_s=100$~MeV (open triangles),
$m_s=200$~MeV (full triangles), $m_s=300$~MeV (stars), respectively.
}
\label{fig1}
\end{figure}
%
In the above, $\rho$ is the total baryon number density in the system ($\rho=A_B/V$),and the color degeneracy and baryon number $1/3$ of the quarks 
have been taken into account.
From the above formulas we calculate the energy per baryon number to be:
\begin{equation}
\frac{E_{tot}}{A_B}=\frac{\epsilon_{tot}}{\rho}.
\label{Eperpart}
\end{equation}
In Fig.~\ref{fig1} the results of the minimal energy per baryon 
(\ref{Eperpart}) corresponding to $B=60, 100, 150$~MeV/fm$^3$ are shown as a
function of $R_s$.
For each value of $B$ we explore three different values of the strange 
mass, $m_s=100, 200, 300$~MeV, and we compare these results with the 
experimental nucleon and hyperon masses (full circles).
We have also evaluated, according to formula (3.6) of Ref.~\cite{DeGrand75},
the baryonic masses which are obtained within the same model employed for
bulk strange matter, using the same sets of bag parameter and strange quark 
mass.
As it appears from the figure, the three lines corresponding to the 
different values of $m_s$ are much lower than the experimental hyperon masses 
for $B=60$~MeV/fm$^3$ and $B=100$~MeV/fm$^3$, while this is not the case for 
$B=150$~MeV/fm$^3$ and $m_s$=300 MeV; however, if we compare the energy of 
strange matter with the corresponding theoretical masses of the various 
hyperons, we find that strange matter is {\it always} lower in energy, and
thus more stable. We can therefore conclude that the MIT bag model without 
perturbative gluon corrections allows the existence of strangelets.
\begin{figure}
\includegraphics[width=\textwidth]{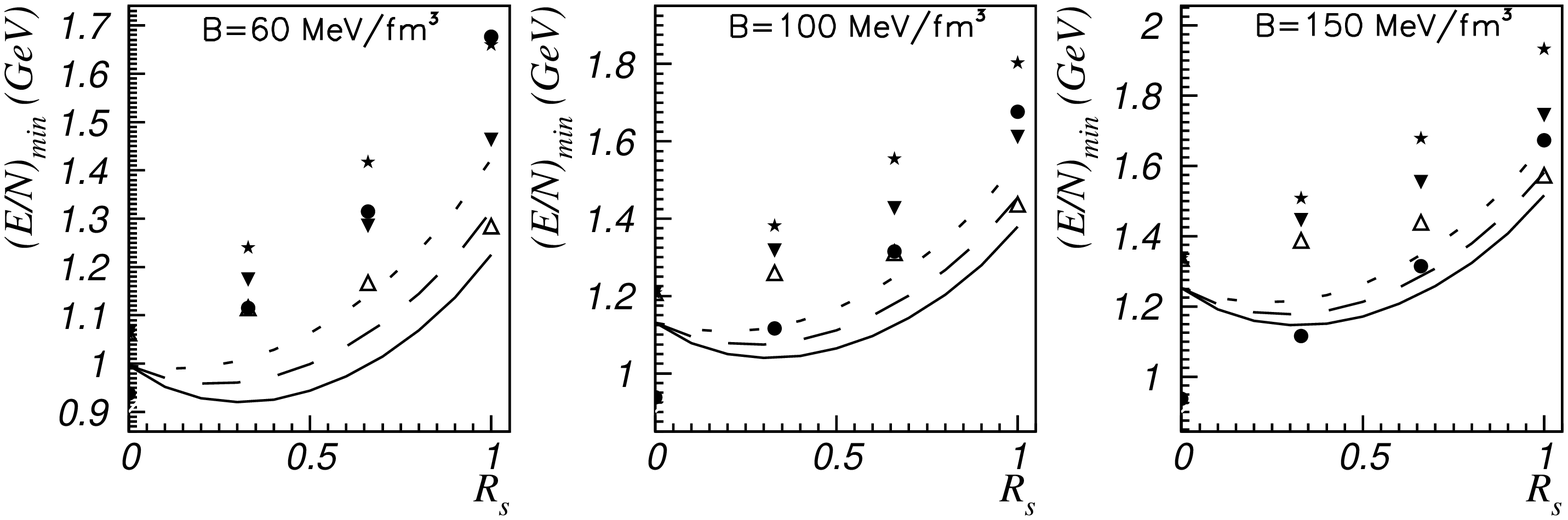}
\caption{Minimal energy per baryon number in the MIT bag model, including the 
OGE potential with $\alpha_s=0.5$, as a function of the strangeness fraction 
$R_s=\rho_s/\rho$. The continuous line corresponds to $m_s=100$~MeV, the 
dashed line to $m_s=200$~MeV and the dotted line to $m_s=300$~MeV. 
Full circles represent the experimental masses, the other points refer to the masses evaluated in the model, with $m_s=100$~MeV (open triangles), $m_s=200$~MeV (full triangles), $m_s=300$~MeV (stars), respectively.
}
\label{fig2}
\end{figure}
\subsection{MIT bag model with perturbative gluons}
\begin{figure}
\includegraphics[width=\textwidth]{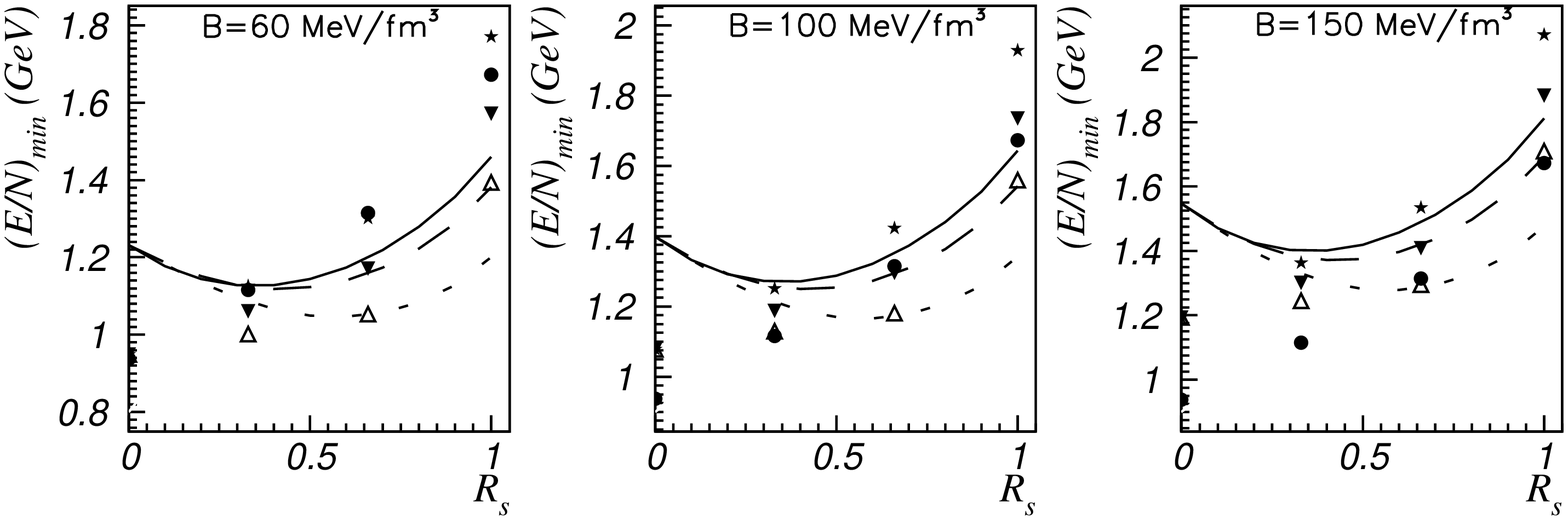}
\caption{The same as in Fig.~\ref{fig2}, but for $\alpha_s=2.2$.
}
\label{fig3}
\end{figure}

We consider now the effects of introducing in the calculation perturbative 
corrections due to the exchange of gluons . At first order in $\alpha_s$, 
two contributions to the energy can be considered, the direct and the exchange
one. Since the system is globally colorless the direct term vanishes, while 
the exchange one gives the following contribution to the energy density of 
quarks of flavor $f$ ~\cite{fahri84}:
\begin{eqnarray}
\epsilon_f^{OGE}&&=-\frac{\alpha_s}{\pi^3}m_f^4\left\{x_f^4-\frac{3}{2}
\left[\ln\left(\frac{x_f+\eta_f}{\eta_f}\right)-x_f\eta_f\right]^2+\right.
\nonumber\\
&&\qquad \left. +
\frac{3}{2}\ln^2\left(\frac{1}{\eta_f}\right)
-3\ln\left(\frac{\mu}{m_f\eta_f}\right)\left[\eta_fx_f
-\ln\left(x_f+\eta_f\right)\right]\right\}\,.
\label{ogemit}
\end{eqnarray}
Here:
\begin{eqnarray}
x_f &=& \frac{k_{F_f}}{m_f}
\nonumber\\
\eta_f &=& \sqrt{1+x_f^2}.
\nonumber
\end{eqnarray}
and $\mu$ is a renormalization scale, for  which we choose the value
$\mu=313$~MeV, according to Ref.~\cite{fahri84}.
For sake of illustration, we adopted two different values for $\alpha_s$, 
a small perturbative value ($\alpha_s=0.5$), which is in line with the 
choices and motivations of Fahri and Jaffe~\cite{fahri84}, and the 
canonical value which was employed  by DeGrand {\it et al.}~\cite{DeGrand75} 
($\alpha_s=2.2$), to reproduce the hyperon masses.
The corresponding results are illustrated in Figs.~\ref{fig2} and \ref{fig3}.
From Fig.~\ref{fig2} we can see that, even after the inclusion of 
perturbative gluons, strangelets are more stable than hyperons
for almost all values of the model parameters. However, when we 
use the stronger coupling of Fig.~\ref{fig3} the stability of strange
matter (and hence strangelets) becomes questionable, particularly 
for low values of the strange mass $m_s$. Only for $m_s=300$~MeV the
theoretical masses of hyperons always lie above the energy of bulk matter
(not so the experimental masses).

From this analysis we can conclude (in agreement with previous findings)
that, apart from rather extreme choices of the model parameters, metastable 
strangelets can exist in the MIT bag model.

\section{Strangelets in the Color Dielectric Model}

The Color Dielectric Model provides absolute confinement of quarks through
their interaction with a scalar field $\chi$ which represents a 
multi--gluon state and produces a density dependent constituent mass 
(see for example the review 
articles~\cite{wilets,birse90,pirner92}) 

The typical Lagrangian of the CDM reads:
\begin{eqnarray}
{\cal L} &&= \sum_{f=u,d,s}{\bar{\psi_f}} i\gamma^{\mu}\left(\partial_{\mu}
-ig_s\frac{\lambda^a}{2}A^a_{\mu}\right)\psi_f
-\frac{g f_{\pi}}{\chi}\sum_{f=u,d}{\bar{\psi_f}} \psi_f
-m_s\left(\chi\right) {\bar{\psi_s}} \psi_s +
\nonumber\\
&&\quad
+\frac{1}{2}\left(\partial_{\mu}\chi\right)^2-U\left(\chi\right)
-\frac{1}{4}\kappa\left(\chi\right)F^a_{\mu\nu}F^{a \mu\nu}\,,
\label{CDMlagr}
\end{eqnarray}
where $\psi_f$ are the quark fields, $A^a_{\mu}$ is the (effective) gluon 
field, $F^a_{\mu\nu}$ its strength tensor and $\chi$ is the color dielectric 
field; $g_s$ is the strong (colour) coupling ($g_s^2/4\pi=\alpha_s$). 

The $u$ and $d$ quark mass terms arise as a consequence of their interaction 
with the $\chi$--field and read:
\begin{equation}
m_{u,d}=\frac{g f_{\pi}}{\chi},
\label{udmass}
\end{equation}
where $g$ is a parameter of the model and $f_\pi$ the pion decay constant, 
which is fixed to its experimental value, $f_\pi=93$~MeV.
 For the strange quark mass we consider two different 
versions of the 3--flavors CDM, namely a {\it scaling model}, with
\begin{equation}
 m_s= \frac{g' f_{\pi}}{\chi}\,,
\label{smass1}
\end{equation}
and a {\it non--scaling model}, with a constant shift of the $s$--mass
with respect to the $u,d$--one:
\begin{equation}
 m_s=\frac{g f_{\pi}}{\chi} +\Delta m \equiv m_{u,d}+\Delta m \,.
\label{smass2}
\end{equation}
In the above $g'$ (or $\Delta m$) is another parameter of the model.

Concerning the color dielectric field, there exist in the literature several 
options, both 
for its coupling to the gluon tensor and for the potential $U(\chi)$. 
We adopt here both the single minimum (SM), quadratic potential:
\begin{equation}
U_{SM}\left(\chi\right)=\frac{1}{2}M^2\chi^2\,,
\label{sm}
\end{equation}
which introduces the third parameter of the model, $M$ (the mass of the 
glueball), and the double minimum (DM), quartic potential:
\begin{equation}
U_{DM}\left(\chi\right)=\left(\frac{1}{2}\frac{M^2}{\chi_0^2}-\frac{3B}
{\chi_0^4}\right)\chi^4+\left(\frac{4B}{\chi_0^3}-\frac{M^2}{\chi_0}\right)
\chi^3+\frac{1}{2}M^2\chi^2.
\label{dm}
\end{equation}
The latter introduces an extra parameter, the bag pressure B, while the 
parameter $\chi_0$ is used to make the ratio $\chi/\chi_0$ dimensionless.
 The color--dielectric function, $\kappa(\chi)$, is usually 
assumed to be a quadratic or quartic function of $\chi$: we will use both
options and hence we set:
\begin{equation}
\kappa\left(\chi\right)=\left(\frac{\chi}{\chi_0}\right)^{\beta},
\qquad\qquad{\mathrm{ with}}\qquad \beta=2,4\ .
\label{cappa}
\end{equation}

The field equations are solved in the mean field approximation and 
neglecting the gluon fields: the latter are subsequently taken into 
account as a perturbation.
The unperturbed (i.e. without gluon contribution) energy density reads:
\begin{eqnarray}
\epsilon_0=&&\sum_{f=u,d,s}\frac{3}{8\pi^2}\left\{ m_f^4\ln\left(\frac{m_f}
{k_{F_f}+\sqrt{k_{F_f}^2+m_f^2}}\right)\right.
\nonumber\\
&& \left.
+ k_{F_f}\sqrt{k_{F_f}^2+m_f^2}\left(2k_{F_f}^2+m_f^2\right)\right\}
+U\left(\bar{\chi}\right)\,,
\label{eps0f}
\end{eqnarray}
the quark masses being given by eqs.~(\ref{udmass}) and (\ref{smass1}) 
[or (\ref{smass2})] with $\chi=\bar{\chi}$.

Beyond $\epsilon_0$ we have perturbatively taken 
into account, to order $\alpha_s$, the exchange of gluons, whose contribution
to the energy density of an infinite, color singlet system is the analogous of 
eq.~(\ref{ogemit}), but with the quark masses defined by Eqs.~(\ref{udmass}) 
and (\ref{smass1}) [or (\ref{smass2})], and with an effective strong coupling 
constant (dressed by the colour dielectric function), which reads:
\begin{equation}
\widetilde\alpha_s = \alpha_s\left(\frac{\chi_0}{\bar{\chi}}\right)^{\beta}
\label{alfaeff1}
\end{equation}
as it can be deduced from the model Lagrangian.
 Eq.~(\ref{ogemit}) only 
contains the exchange term of OGE, the direct one vanishing for infinite 
quark matter: at small baryonic densities the attractive electric 
contribution dominates the energy density; on the contrary the repulsive 
magnetic contribution becomes the dominant one at large densities.

Indeed the divergent behavior of the electric term  for $\rho\to 0$ could
prevent  a perturbative treatment of OGE in this regime. We have 
overcome this difficulty by taking into account the Debye screening of the 
gluon propagator in the presence of a polarized medium. This can be 
achieved by replacing (\ref{alfaeff1}) with a new effective coupling:
\begin{equation}
\label{alfaeff2}
\alpha_s^{eff}(q) = \widetilde\alpha_s\frac{q^2}
{q^2+\frac{1}{2}\sum_{f=u,d,s} 16\widetilde\alpha_s m_f k_{F_f}^2
\Pi(q/k_{F_f})}\,,
\end{equation}
$\Pi(y)$ being the static limit of the polarization propagator~\cite{FetWal}:
\begin{equation}
\Pi\left(y\right)=\frac{1}{2}-\frac{1}{2y}\left(1-\frac{1}{4}y^2\right)
\ln\left|\frac{1-\frac{1}{2}y}{1+\frac{1}{2}y}\right|\,.
\label{gpol}
\end{equation}

Actually this expression should be utilized in the momentum dependent 
$V_{OGE}(\vec q)$ and then integrated to obtain the new expression for
$\epsilon_f^{OGE}$. For simplicity, since the 
$q$--integration is extended only up 
to $k_{F_f}$ and the function $\Pi(y)$ varies at most by $9\%$ in the range
$0\le y \le 1$, we have adopted $q=k_{F_f}$.\\

\subsection{Stability of strangelets in the CDM: I}

We consider here the work by Aoki {\it et al.}~\cite{aoki91}: these authors
solve self-consistently the mean field equations for quarks, color dielectric 
field and gluons, starting from a CDM Lagrangian with the Double Minimum 
potential (\ref{dm}) for the color dielectric field. This model is known to 
produce an unrealistic, too large binding energy in infinite quark 
matter~\cite{Drago95,Barone95}; concerning the color dielectric function, 
Aoki {\it et al.} choose $\beta=2$; they employ both the scaling and 
non--scaling version of the model, with two different sets for the model 
parameters whose values are dictated by two different and extreme choices 
for the ``bag'' parameter $B$: 
$B^{1/4}=0$~MeV, with two degenerate vacua, and a large bag pressure, 
$B^{1/4}=103.5$~MeV. The latter value of B is chosen to be as large as 
possible, but with the requirement that the two-phase picture must hold 
inside hadrons. In their calculation only the strange quark mass has to
be considered as a truly free parameter, the remaining ones having been fixed 
in a previous work on the non--strange baryons~\cite{aoki90}.

We evaluate the minimum energy per baryon number using cases B and D 
(corresponding to $B\ne 0$) of the work of Aoki {\it et al.}, both 
without and with the perturbative exchange of a gluon.
\begin{figure}
\includegraphics[width=\textwidth]{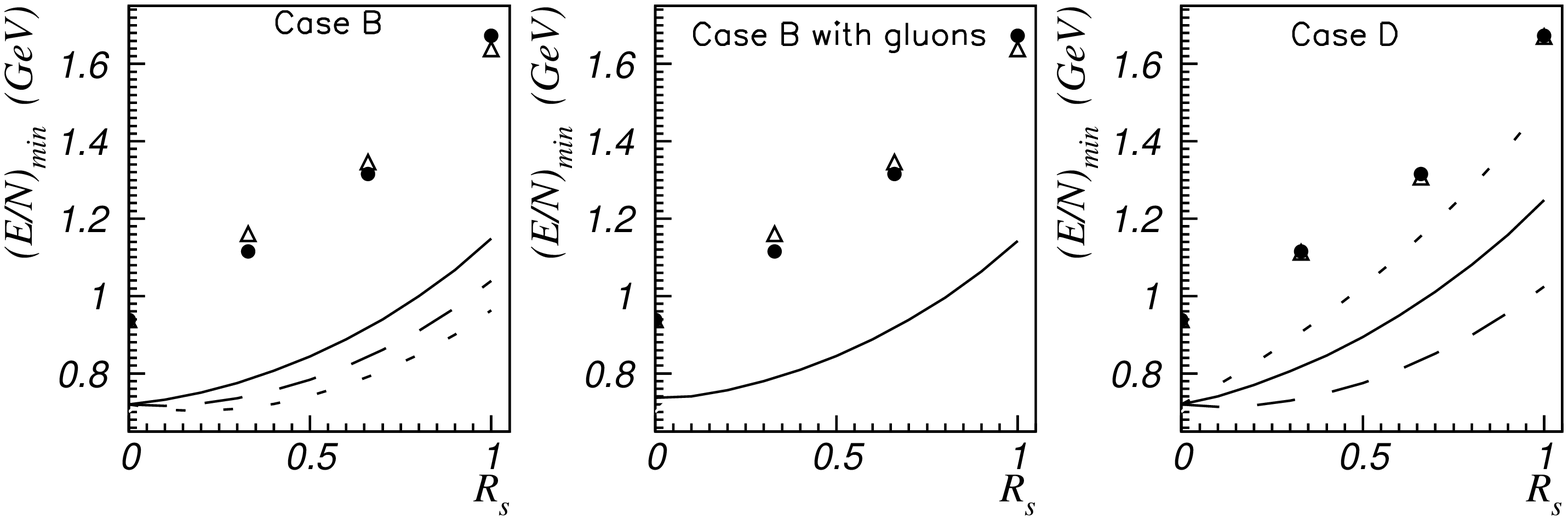}
\caption{Minimal energy per baryon number in the CDM, as a function of 
$R_s=\rho_s/\rho$, for the cases B (with and without gluons) and D 
(solid lines). 
Full circles are the experimental hyperon masses, while triangular dots are 
the masses calculated in Ref.~\cite{aoki91}. In the first panel
 the curves corresponding to $g'=106.6$~MeV (dashed line) and $g'=85.7$~MeV 
(dotted line) are also presented, while 
in the third panel the curves corresponding to 
$\Delta m=312$~MeV (dotted line) and $\Delta m=112$~MeV (dashed line) 
are shown. The remaining parameters of the cases B and D, respectively, 
are kept unaltered.
}
\label{fig4}
\end{figure}

As we can see from Fig.~\ref{fig4}, this version of the 
CDM seems to favour strangelets as a (meta)--stable form of matter. 
This is due to the fact that when the DM potential is used to study hadrons, 
i.e. confined objects, a large contribution to the hadronic mass is given by 
the space fluctuations of the fields. When this version of the model is used 
to describe infinite quark matter, these contributions vanish due to the 
homogeneity of the system. 
 For this reason, deconfined matter is favoured in 
this version of the model, which would even imply spontaneous decay of 
ordinary nuclei or two flavor nuclear matter into quark matter.\\ 
The effect of perturbative gluons in this model is very small, due to the 
rather strong Debye screening, which we have included.  Whether or not we 
take into account gluon corrections, strange matter always appears to be more
stable than baryons. 

\subsection{Stability of strangelets in the CDM: II}

In this subsection we follow the approach of 
J.~McGovern~\cite{McGovern}, using the model Lagrangian reported 
in eq.~(\ref{CDMlagr}) with $\beta=4$ in the color dielectric function.
McGovern employs only the scaling model and the Single Minimum potential, 
with different values of the parameters. In this case 
the behavior of $\alpha_s^{eff}(\bar\chi)$ is even more divergent, for small 
densities, than in the case $\beta=2$ previously considered: hence the use 
of Debye screening in the effective strong coupling constant is mandatory.
In Ref.~\cite{McGovern} two different sets for the model parameters are used:
they allow to satisfactorily reproduce the splittings between hyperon masses, 
but the absolute values of the masses themselves are generally too large.
\begin{figure}
\includegraphics[width=\textwidth]{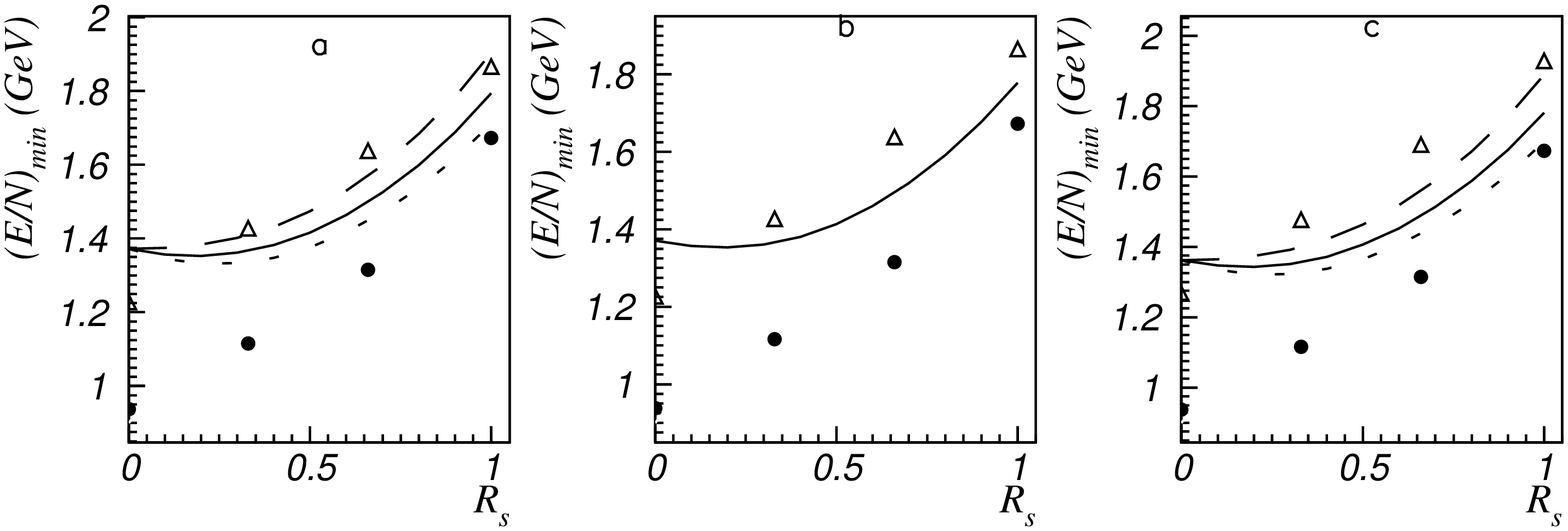}
\caption{Minimal energy per baryon number as a function of $R_s=\rho_s/\rho$ 
for the Single Minimum version of the CDM.
The various panels correspond to: (a) parameter set I without gluons, 
(b) parameter set I with gluons, (c) parameter set II 
without gluons. Full circles are the 
experimental baryon masses, while triangular dots are the masses calculated 
in Ref.~\cite{McGovern}. In the first and third panels the calculations 
obtained with
$g'/g=1.89$ (long-dashed lines) and $g'/g=1.37$ (short-dashed lines) are also
shown.}
\label{fig5}
\end{figure} 
In Fig. \ref{fig5} we show our results, comparing our curves with both the 
experimental and the theoretical masses. As we can see, also in this case the 
inclusion of perturbative gluons is rather irrelevant. The curves 
corresponding to strange matter are well above the experimental masses, and 
below the theoretical ones. Yet, if we take into account surface effects, 
which would increase our curves of about  $50\div100$ MeV, only for 
$R_s\simeq\frac{2}{3}$ strangelets are 
(marginally) allowed by the present calculation and a more refined one, 
taking into account surface energy contributions, is needed to clarify the 
situation. We notice that a larger strange quark mass ($g'/g=1.89$) obviously
excludes stable strangelets, while the smaller $m_s$ value ($g'/g=1.37$)
does not substantially alter the above considerations.

\section{Conclusions}
The aim of this contribution was to compare the predictions about strangelet 
stability within the MIT bag model and the Color Dielectric Model, and to draw 
conclusions about their model dependence: we have compared the curves 
corresponding to the minimum energy per baryon number to the mass of hyperons 
having the same strangeness fraction and calculated within the same model and 
parameter values that we adopt in our calculations.

The analysis shows that the existence of (stable) strangelets
is supported only by those models which entail a two--phase picture of 
hadrons, namely which maintain a false vacuum inside hadrons.
This happens both in the MIT bag model, and in the Double Minimum version
of the Color Dielectric Model. The Single Minimum version of the CDM does not 
allow the existence of strangelets, independently of the parameter sets used
to perform the calculations. 

The conclusions that we can draw indicate that the stability of strangelets 
depends rather crucially on the model employed; this fact can set serious 
challenges to the search for strangelets in heavy ion collisions.

\end{document}